\documentclass[aps,twocolumn,preprintnumbers,superscriptaddress,amsmath,amssymb,footinbib,prl]{revtex4}
\usepackage{epsfig}
\usepackage{graphicx}
\usepackage{bm}
\newcommand{\be}{\begin{equation}}
\newcommand{\ee}{\end{equation}}
\newcommand{\bea}{\begin{eqnarray}}
\newcommand{\eea}{\end{eqnarray}}

\begin{document}
\title{(3+1)-Dimensional Hydrodynamic Expansion with a Critical Point from Realistic Initial Conditions}

\author{J.~Steinheimer}
\affiliation{Institut f\"ur Theoretische Physik, Johann Wolfgang Goethe-Universit\"at, Max-von-Laue-Str.~1, 
D-60438 Frankfurt am Main, Germany}

\author{M.~Bleicher}
\affiliation{Institut f\"ur Theoretische Physik, Johann Wolfgang Goethe-Universit\"at, Max-von-Laue-Str.~1, 
D-60438 Frankfurt am Main, Germany}

\author{H.~Petersen}
\affiliation{Institut f\"ur Theoretische Physik, Johann Wolfgang Goethe-Universit\"at, Max-von-Laue-Str.~1, 
D-60438 Frankfurt am Main, Germany}
\affiliation{Frankfurt Institute for Advanced Studies (FIAS), Ruth-Moufang-Str.~1, D-60438 Frankfurt am Main,
Germany}

\author{S.~Schramm}
\affiliation{Institut f\"ur Theoretische Physik, Johann Wolfgang Goethe-Universit\"at, Max-von-Laue-Str.~1, 
D-60438 Frankfurt am Main, Germany}

\author{H.~St\"ocker}
\affiliation{Institut f\"ur Theoretische Physik, Johann Wolfgang Goethe-Universit\"at, Max-von-Laue-Str.~1, 
D-60438 Frankfurt am Main, Germany} 
\affiliation{Frankfurt Institute for Advanced Studies (FIAS), Ruth-Moufang-Str.~1, D-60438 Frankfurt am Main,
Germany}
\affiliation{Gesellschaft f\"ur Schwerionenforschung (GSI), Planckstr.~1, D-64291 Darmstadt, Germany}

\author{D.~Zschiesche}
\affiliation{Institut f\"ur Theoretische Physik, Johann Wolfgang Goethe-Universit\"at, Max-von-Laue-Str.~1, 
D-60438 Frankfurt am Main, Germany} 

\begin{abstract}
We investigate a (3+1)-dimensional hydrodynamic expansion of the hot and dense system
created in head-on collisions of Pb+Pb/Au+Au at beam energies from $5-200A~$GeV. An equation of
state that incorporates a critical end point (CEP) in line with the lattice data is used. The necessary
initial conditions for the hydrodynamic evolution are taken from a microscopic transport
approach (UrQMD). We compare the properties of the initial state and the full hydrodynamical
calculation with an isentropic expansion employing an initial state from a simple overlap model.
We find that the specific entropy ($S/A$) from both initial conditions is very similar and
only depends on the underlying equation of state. Using the chiral (hadronic) equation of
state we investigate the expansion paths for both initial conditions. Defining a critical
area around the critical point, we show at what beam energies one can expect to have a
sizable fraction of the system  close to the critical point. Finally, we emphasise the
importance of the equation of state of strongly interacting matter, in the (experimental)
search for the CEP.
\end{abstract}

\maketitle

\noindent
Heavy ion collisions at intermediate incident beam energies ($5-200A~$GeV) have
recently attracted more and more attention, since one expects to be able to scan a wide - and
by current expectations highly interesting \cite{Heinz:2000bk,Gyulassy:2004zy} - region of
temperature $T$ and baryo-chemical potential $\mu_B$ in the QCD phase diagram. Following recent 
theoretical investigations (for
recent lattice QCD results see \cite{Fodor:2001pe,Fodor:2007vv,Karsch:2004wd}, for
phenomenological studies see
\cite{Stephanov:1998dy,Gazdzicki:1998vd,Stephanov:1999zu,Bravina:1999dh,Bravina:2000dk,Gazdzicki:2004ef,Arsene:2006vf}),
one hopes to find experimental evidence for a phase transition from hadronic matter (with
broken chiral symmetry), to a Quark-Gluon Plasma (QGP) phase (where chiral symmetry is restored).
Especially the so called critical end point (CEP), a point in the phase diagram that terminates the
phase transition-line of the first order transition (which is expected for high chemical
potentials), is of great interest. 

Key observables like the directed and elliptic flow $v_1$ and $v_2$
\cite{Hofmann:1976dy,Stoecker:1986ci,Sorge:1998mk,Ollitrault:1992bk,Hung:1994eq,Rischke:1996nq,Sorge:1996pc,Heiselberg:1998es,Soff:1999yg,Brachmann:1999xt,Csernai:1999nf,Zhang:1999rs,Kolb:2000sd,Bleicher:2000sx,Kolb:2003dz,Stoecker:2004qu,Zhu:2005qa,Petersen:2006vm} - but also particle multiplicities, ratios and their fluctuations - are of imminent
interest. They have been predicted and sometimes already shown to be sensitive to the
properties of the QCD matter, i.e. to the Equation of State (EoS) and the active degrees of
freedom, in the early stage of the reaction. And indeed, the energy dependences of various
observables show anomalies at low SPS energies which might be related to the onset of
deconfinement \cite{Gazdzicki:2004ef,Gazdzicki:1998vd}.

For a hydrodynamical modelling of heavy ion reactions to study these observables one needs to specify 
initial conditions, i.e. an infinite set of  space-time points with their corresponding energy- and baryon density.
Since experimental data provides mainly information that is integrated over the systems time evolution, the initial 
state for hydrodynamical simulations has to be inferred from model assumptions or by educated 'guessing' in 
comparison to data. The latter approach - usually applied for relativistic hydrodynamical simulations of nuclear 
collisions - is, however, by no means straight forward and highly non-trivial: the connections between (observed) 
final state and the inferred  initial conditions is blurred by the unknown equation of state, potential viscosity effects,
and freeze-out problems. Another issue concerns the assumption of thermal equilibrium, which is probably not true at 
least for the early stage of a heavy ion collisions at intermediate energies. 

There have been attempts to solve these problems by describing such collisions with viscous or multi-fluid-hydrodynamic 
models 
\cite{Mishustin:1988mj,Mishustin:1989nj,Katscher:1993xs,Brachmann:1997bq,Reiter:1998uq,Bleicher:1998xi,Brachmann:1999xt,Brachmann:1999mp,Dumitru:2000ai,Russkikh:2003ma,Ivanov:2005yw,Toneev:2005yy,Baier:2006gy,Song:2007fn}, 
but the practical application of these models is difficult. To avoid (some of) these problems in 
this paper, we describe the initial stages of the collision with a non-equilibrium
transport model (UrQMD). We then use the so obtained distributions for energy- and baryon-density as 
initial conditions for a one-fluid but fully (3+1)-dimensional hydrodynamical calculation. 
For the hydrodynamical evolution an EoS with a first order chiral phase transition and
a CEP at finite $\mu_B$ is applied.

This paper is organised as follows: In the first part we describe the chiral equation of state that is used in more detail and explain how the structure of the phase diagram with a critical end point is modelled. Afterwards two different scenarios for the initial conditions and the subsequent evolution are compared. The full hydro evolution with microscopic initial conditions is contrasted with lines of constant entropy per baryon number from a simple overlap model. Then, the results concerning the CEP are presented and the influence of the equation of state on the evolution is studied. The last part summarises the paper. 

Let us start with the discussion of the equation of state. The present chiral hadronic $SU(3)$ Lagrangian incorporates the complete set of
baryons from the lowest flavour-$SU(3)$ octet, as well as
the entire multiplets of scalar, pseudo-scalar, vector and axial-vector
mesons \cite{Papazoglou:1998vr}. In mean-field approximation, the
expectation values of the scalar fields relevant for symmetric nuclear
matter correspond to the non-strange and strange chiral quark
condensates, namely the $\sigma$ and its $s\bar{s}$ counterpart
$\zeta$, respectively, and further the $\omega$ and $\phi$ vector
meson fields. Another scalar iso-scalar field, the dilaton $\chi$, is
introduced to model the QCD scale anomaly. However, if $\chi$
does not couple strongly to baryonic degrees of freedom it remains
essentially ``frozen'' below the chiral transition \cite{Papazoglou:1998vr}.
Consequently, we focus here on the role of the
quark condensates.

Interactions between baryons and scalar (BM) or vector (BV)
mesons, respectively, are introduced as
\begin{eqnarray}
\label{L_BM+V}
{\cal L}_{\rm BM} &=&
-\sum_{i}   \overline{\psi}_i \left( g_{i\sigma}\sigma + g_{i\zeta}\zeta
\right) \psi_i
~,\\
{\cal L}_{\rm BV}
  &=& -\sum_{i}   \overline{\psi}_i
\left( g_{i\omega}\gamma_0\omega^0 + g_{i \phi}\gamma_0 \phi^0 \right) \psi_i
~,
\end{eqnarray}
Here, $i$ sums over the baryon octet ($N$, $\Lambda$, $\Sigma$, $\Xi$).
A term ${\cal L}_{\rm vec}$ with
mass terms and quartic self-interaction of the vector mesons is also added:
\begin{eqnarray}
{\cal L}_{\rm vec} &=& \frac{1}{2}
a_\omega \chi^2 \omega^2 + \frac{1}{2}
a_\phi \chi^2 \phi^2
+ g_4^{\,4} (\omega^4 + 2 \phi^4 )~. \nonumber
\end{eqnarray}
The scalar self-interactions are
\begin{eqnarray}
\label{L_0}
{\cal L}_0 &=& -\frac{1}{2} k_0 \chi^2 (\sigma^2+\zeta^2) + k_1
    (\sigma^2+\zeta^2)^2 + k_2 ( \frac{ \sigma^4}{2} + \zeta^4)
    \nonumber \\ & &{} + k_3 \chi \sigma^2 \zeta
 - k_4 \chi^4 - \frac{1}{4}\chi^4  \ln\frac{\chi^4}{\chi_0^{\,4}}
   +\frac{\delta}{3} \chi^4 \ln\frac{\sigma^2\zeta}{\sigma_0^{\,2} \zeta_0}
~.
\end{eqnarray}
Interactions between the scalar mesons induce the spontaneous
breaking of chiral symmetry (first line) and the scale breaking via
the dilaton field $\chi$ (last two terms).

Non-zero current quark masses break chiral symmetry explicitly in
QCD. In the effective Lagrangian this corresponds to terms such as
\begin{eqnarray}
{\cal L}_{\rm SB} &=& -\frac{\chi^2}{\chi_0^{\,2}}
\left[m_\pi^2 f_\pi \sigma + (\sqrt{2}m_K^2 f_K -
\frac{1}{\sqrt{2}}
m_{\pi}^2 f_{\pi})\zeta \right]~.
\end{eqnarray}

According to ${\cal L}_{\rm BM}$ (\ref{L_BM+V}), the effective
masses of the baryons,
$m_i^*(\sigma,\zeta)=g_{i\sigma}\,\sigma+g_{i\zeta}\,\zeta$\,,
are generated through their coupling to the chiral condensates,
which attain non-zero vacuum expectation values due to their
self-interactions \cite{Papazoglou:1998vr} in ${\cal L}_0$ (\ref{L_0}).
The effective masses of the mesons are obtained as the second
derivatives of the mesonic potential
${\cal V}_{\rm Meson}
 \equiv
 - {\cal L}_0 - {\cal L}_{\rm vec} - {\cal L}_{\rm SB}
$
about its minimum.

The baryon-vector couplings $g_{i\omega}$ and $g_{i\phi}$
result from pure $f$-type
coupling as discussed in \cite{Papazoglou:1998vr},
$g_{i\omega} = (n^i_q-n^i_{\bar{q}}) g_{8}^V$\,,
$g_{i\phi}   = -(n^i_s-n^i_{\bar{s}}) \sqrt{2} g_{8}^V$\,,
where 
$g_8^V$ denotes the vector coupling of the baryon
octet and
$n^i$ the number of constituent quarks of species $i$ in a given hadron.
The resulting relative couplings agree with additive quark model
constraints.

All parameters of the model discussed so far are fixed by either
symmetry relations, hadronic vacuum observables or nuclear matter
saturation properties (for details see \cite{Papazoglou:1998vr}).
In addition, the model also provides a satisfactory description of realistic
(finite-size and isospin asymmetric) nuclei and of neutron stars
\cite{Papazoglou:1998vr,nucl-th/0210053,nucl-th/0207060}.
%

If the baryonic degrees of freedom are restricted to the members
of the lowest lying octet,
the model exhibits a smooth decrease of the chiral condensates
(crossover)
for both high $T$ and high $\mu$  \cite{Papazoglou:1998vr,nucl-th/0107037}.
However, additional baryonic
degrees of freedom
may change this into a first-order phase transition
in certain regimes of the $T$-$\mu_q$ plane, depending on the
couplings \cite{Theis:1984qc,nucl-th/0107037,nucl-th/0407117}.
To model the influence of such heavy baryonic states, we add a single resonance
with mass
$m_R = m_{0} + g_R \sigma$
and vector coupling
$g_{R\omega} = r_V g_{N\omega}$\,.
The mass parameters, $m_{0}$\,, $g_R$  and the relative
vector coupling $r_V$ represent free parameters, adjusted
to reproduce the phase diagram discussed
above\footnote{Note that instead of an explicit mass term we could have
coupled the resonance to the dilaton $\chi$\,.}. In principle, of course,
one should couple the entire spectrum of resonances to the scalar and
vector fields. However, to keep the number of additional
couplings small, we effectively describe the couplings of all higher
baryonic resonances by a single state with adjustable couplings, mass and degeneracy.
This method of replacing the influence of many states by an effective resonance
has a long tradition in scattering theory and related fields (see e.g. \cite{Mott}).

In what follows, the meson fields are replaced by their (classical)
expectation values, which corresponds to neglecting quantum and
thermal fluctuations. Fermions have to be integrated out
to one-loop. The grand canonical potential can then be written as
\begin{eqnarray}
\label{thermpot}
   {\Omega}/{V}&=& -{\cal L}_{\rm vec} - {\cal L}_0 - {\cal L}_{\rm SB}
-{\cal V}_{\rm vac} \\
& &{} -T \sum_{i \in B} \frac{\gamma_i }{(2 \pi)^3} 
\int d^3k \left[\ln{\left(1 + e^{-\frac 1T[E^{\ast}_i(k)-\mu^{\ast}_i]}\right)} \right]
\nonumber \\
& &{}+T \sum_{l\in M} \frac{\gamma_l}{(2 \pi)^3} 
\int d^3k \left[\ln{\left(1 - e^{-\frac 1T[E^{\ast}_l(k)-\mu^{\ast}_l]}\right)
}\right], \nonumber
\end{eqnarray}
where $\gamma_B, \gamma_M$ denote the baryonic and mesonic
spin-isospin degeneracy factors and $E^{\ast}_{B,M} (k)
= \sqrt{{k}^2+{m_{B,M}^*}^2}$ are the corresponding single
particle energies. In the second line we sum over the baryon octet
states plus the additional heavy resonance with degeneracy $\gamma_R$
(assumed to be 16). The effective baryo-chemical potentials are
$\mu^{\ast}_i = \mu_i-g_{i \omega} \omega-g_{i \phi} \phi$, with
$\mu_i= (n^i_q - n^i_{\bar{q}}) \mu_q + (n^i_s - n^i_{\bar{s}})
\mu_s$. The chemical potentials of the mesons are given by the sum of
the corresponding quark and anti-quark chemical potentials.  The
vacuum energy ${\cal V}_{\rm vac}$ (the potential at $\rho_B=T=0$) has
been subtracted.

By extremizing $\Omega/V$ one obtains self-consistent gap equations
for the meson fields. Here, globally non-strange matter is considered and 
$\mu_s$ for any given $T$ and $\mu_q$ is adjusted to obtain a vanishing net strangeness.
The dominant ``condensates'' are then the $\sigma$ and the $\omega$ fields.
There have also been first attempts to model the dynamical evolution of the condensates
themselves instead of 'locking' them at their equilibrium values (see e.g. \cite{Paech:2003fe,Paech:2005cx}).

Let us now turn to the explanation of the initial conditions that have been used.  
Two different ways to describe the initial conditions and the expansion will be discussed. 
In one setup the energy and density distributions obtained from the UrQMD transport simulation are 
mapped to the thermodynamic quantities, which then serve as initial conditions for the (3+1) dimensional 
hydrodynamic evolution. In the second setup initial energy- and baryon-densities
as obtained from a simple overlap model are employed and then paths of
constant entropy per baryon are followed.

In the first scenario, the Ultra-relativistic Quantum Molecular Dynamics Model (UrQMD 1.3) is used to 
calculate the initial state for the hydrodynamical evolution. This has been done to account for the non-equilibrium in the very early stage of the collision. In this configuration the effect of event-by-event fluctuations of the initial state is naturally included. Due to the early transition time to hydrodynamics, only
initial scatterings, i.e. baryon-baryon collisions and string excitations/fragmentations are relevant here.
As many details of the UrQMD model are not relevant for the present initial state calculation
we refer the interested reader for the details \cite{Bass:1998ca,Bleicher:1999xi}
of the UrQMD model and its thermodynamic properties \cite{Bass:1997xw,Belkacem:1998gy,Bravina:1998it}. 
The coupling between the UrQMD initial state and 
the hydrodynamical evolution happens when the two Lorentz-contracted nuclei have passed through each other. 
This assures that (essentially) all initial baryon-baryon scattering have proceeded and that the energy 
deposition has taken place. It should be noted that the present approach is different from using a kinetic
model for the freeze-out procedure \cite{Anderlik:1998et,Dumitru:1999sf,Bass:1999tu,Magas:1999yb,Bass:2000ib,Teaney:2001gc,Teaney:2001av,Nonaka:2006yn} which is not done in the present investigation, but it is in spirit similar to the NeXSPheRIO 
approach \cite{Paiva:1996nv,Aguiar:2001ac,Socolowski:2004hw,Grassi:2005pm,Andrade:2005tx,Andrade:2006yh,Aguiar:2007zz}.
\begin{figure}[t]
\includegraphics[width=9.5cm]{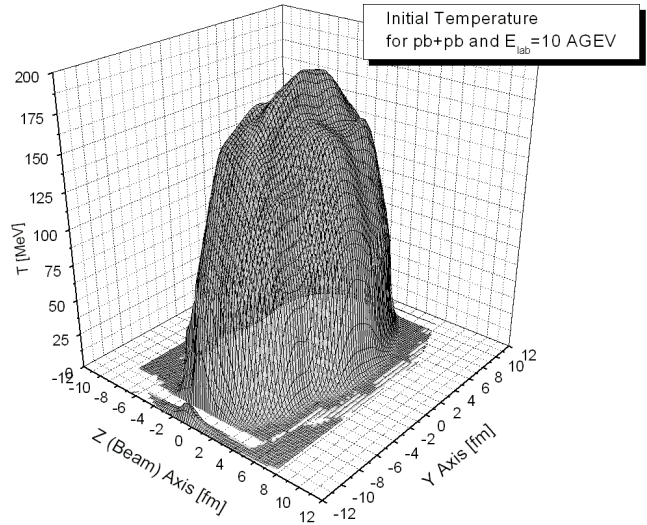}
\caption{\label{Temperature distribution}
The initial temperature distribution in the z-y plane. Where z is the beam-axis and y the out of plane axis.}
\end{figure}
\begin{figure}[h]
\centering
\includegraphics[width=9.0cm]{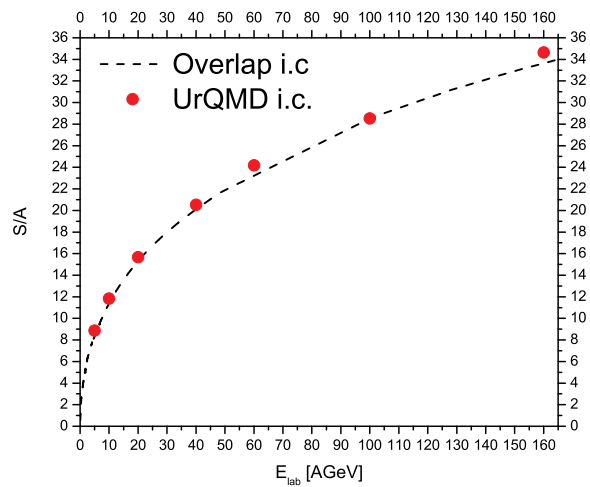}
\caption{\label{sa_ex}
Excitation functions of $S/A$ for the geometrical overlap model and UrQMD initial conditions.}
\end{figure}

To allow for a consistent and numerically stable mapping of the 'point like' particles from UrQMD to the 3-dimensional 
spatial-grid with a cell size of $(0.2  {\rm fm})^3$, each hadron is represented by a Gaussian with a finite width. 
I.e. each particle is described by a three-dimensional Gaussian distribution of its total energy-, momentum- 
(in x-, y-, and z-direction) and baryon number-density. The width of these Gaussians is 
chosen to be $\sigma=1~$fm. A smaller Gaussian widths leads to numerical instabilities (e.g. entropy production) in 
the further hydrodynamical evolution, while a broader width would smear out the initial fluctuations to a large extend.  
To account for the Lorentz-contraction of the nuclei in the longitudinal direction, a gamma-factor (in 
longitudinal direction) is included.
The resulting distribution function, e.g. for the energy density, then reads:
\begin{equation}
\footnotesize{\epsilon_{\rm cf}(x,y,z)=N  \exp{\frac{(x-x_{p})^2+(y-y_{p})^2
+(\gamma_z(z-z_{p}))^2}{2 \sigma^2}}}\quad,
\end{equation}
where $N=(\frac{1}{2 \pi })^{\frac{3}{2}} \frac{\gamma_z}{\sigma^3}  E_{lab} $ provides the proper normalisation,
$\epsilon_{\rm cf} $ and $E_{\rm cf} $ are the energy density and total energy of the 
particle in the computational frame, while $(x_p,y_p,z_p)$ is the position vector of the particle.
Summing over all single particle distribution functions leads to distributions of
 energy-, momentum- and baryon number-densities in each cell.

This is done for Pb+Pb/Au+Au collisions at $E_{\rm lab}=5 - 200A$~GeV with impact parameter $b=0$~fm.
To relate the distributions of energy and baryon number-density to thermodynamic quantities like pressure, 
temperature, chemical potential or entropy-density, the equation of state described above is used.
As an example, Fig. \ref{Temperature distribution} shows the initial temperature distribution 
obtained for $E_{\rm lab} = 10A~$ GeV.

We contrast the microscopically calculated initial conditions described above with a simplified
overlap geometry initial condition. Therefore, we assume that the entire initial beam energy and baryon number
equilibrates in a Lorentz-contracted volume determined by the overlap of projectile and target in the 
center-of-mass frame. This allows to obtain a straightforward estimate for the initial baryon number and energy density:
\begin{eqnarray}
\rho_B^{\rm initial} &= 2\,\gamma_{\rm CMS}\,\rho_0\quad,\\
\epsilon^{\rm initial} &= \sqrt{s}\, \rho_0\, \gamma_{\rm CMS}\quad.
\end{eqnarray}
It was shown in \cite{Reiter:1998uq} that in the energy range of interest here, this rather simple approach
reproduces the specific entropy production from a three-fluid model quite well.

For the subsequent hydrodynamical evolution of the system we apply a fully (3+1)-dimensional one-fluid model.
The hydrodynamical equations are solved by means of the SHASTA (SHarp And Smooth
Transport Algorithm) as described in \cite{Rischke:1995ir}. The EoS with a CEP 
is provided in tabulated form with a fixed step size in energy and baryon density: $\Delta \epsilon = 0.1$ 
and $\Delta n = 0.05$ where $\epsilon$ and $n$ are given in units of nuclear ground state 
densities ($\epsilon_0 \approx 138.5$~MeV and $n_0 \approx 0.15 {\rm fm}^{-3}$). This hydrodynamical model has been 
tested vigorously and applied successfully for various initial conditions and physics investigations \cite{Waldhauser:1992xf,Schneider:1993gd,Rischke:1995ir,Rischke:1995mt,Rischke:1995cm,Gyulassy:1996br,Stoecker:2007su,Betz:2007ie}.
\begin{figure}[t]
\centering
\includegraphics[width=9.0cm]{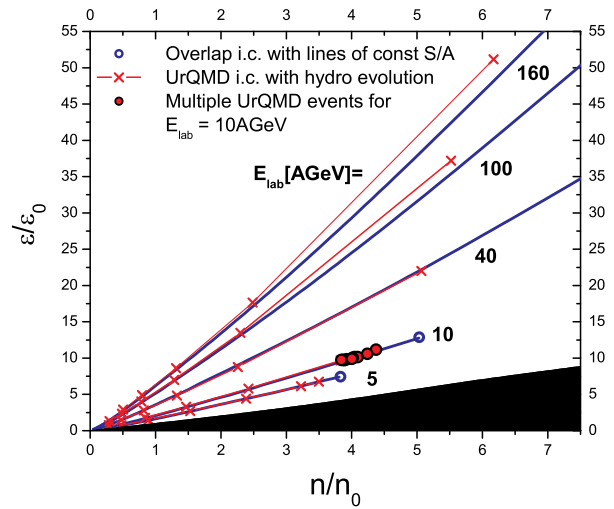}
\caption{\label{erho}
Isentropic expansion paths in units of ground state densities
($\epsilon_0 = 138.5$~MeV and $n_0 = 0.15 {\rm fm}^{-3}$).
Red circles correspond to multiple UrQMD events for the same beam
energy and impact parameter. The black region indicates the region below $T\le0$.}
\end{figure}
\begin{figure}[t]
\centering
\includegraphics[width=9.0cm]{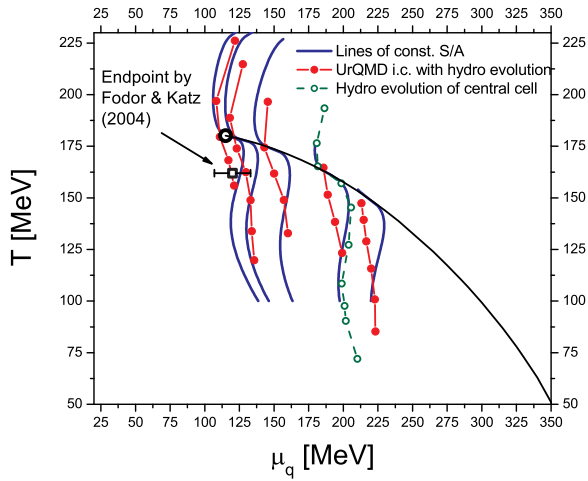}
\caption{\label{tmu}
Isentropic expansion paths in the $T-\mu_q$ plane for very central Pb+Pb/Au+Au reactions.
UrQMD initial conditions with (3+1)-dimensional chiral hydrodynamical evolution (averaged, full red line with 
circles; central cell, dashed green line with circles), isentropic expansion from the overlap model initial 
conditions are shown as full line in blue. Beam energies are from left to right: $160, 100, 40, 10, 5 A$~GeV. 
The phase boundary of the model is shown as full black line with the critical end point, the Fodor and 
Katz critical end point is shown separately with error bars \cite{Fodor:2004nz}.}
\end{figure}

Fig. \ref{sa_ex} depicts the excitation function of the total entropy $S$ per baryon number $A$ for the initial 
stage of the hydrodynamical evolution for both initial conditions (UrQMD, solid circles; overlap model, dashed line). 
We have checked that both quantities are separately conserved throughout the whole hydrodynamical evolution, 
so $S/A$ is a time independent constant\footnote{It should be noted that entropy conservation is numerically 
difficult in the present  approach if the initial conditions are too 'spiky', thus fixing the minimal 
value of the Gaussian width for the smearing of the particles.}. 
Interestingly, the simple geometric overlap model and the UrQMD initial conditions yield basically the 
same value of $S/A $ for a given incident energy. 

The (isentropic) expansion paths for different beam energies in the $\epsilon $- $n $ plane are shown in 
Fig. \ref{erho}. Here $n $ is the baryon-number density.
Again lines of constant $S/A$ for overlap initial conditions (blue open circles and lines), and the 
(3+1)-dimensional hydrodynamical evolution with UrQMD initial conditions are compared.
The mean values are obtained by weighting the value of a specific quantity in a given cell with 
the energy density of that cell. E.g., the mean number density is calculated from:
\begin{equation}
<n> =\frac{\sum_{i,j,k}{ n_{i,j,k}\cdot\epsilon_{i,j,k} } }{\sum_{i,j,k}{\epsilon_{i,j,k}}},
\end{equation}
where $i,j,k $ represent the cell indices.
The mean values for $\epsilon$, $T$ and $\mu_q$ (quark chemical potential) are calculated accordingly.
This has been done for equal time intervals of $\Delta t=2.4$~fm/c.
As expected, for a given beam energy, the dynamical paths obtained from the (3+1)-dimensional hydro 
evolution agree quite well with the lines of constant $S/A$.
The variation in energy and baryon density due to the variation of the initial state in UrQMD even for a fixed
impact parameter and fixed beam energy are studied for $E_{\rm lab}=10A$~GeV and indicated by the red full circles. 

As a next step the hydrodynamic evolution of the system is shown in the $T-\mu_q$
plane in Fig. \ref{tmu}. Also indicated is the first order phase transition
line and the CEP of the employed chiral EoS.
Included are again lines of constant $S/A$ and hydrodynamical evolution paths 
for the same beam energies (from left to right: $160, 100, 40, 10, 5 A$~GeV) 
as in Fig. \ref{erho}. As one can see, the mean temperatures and chemical potentials of the
hydrodynamical evolution are not identical to the respective lines of constant entropy. This is due to 
the averaging procedure while a single cell does follow the isentropic path. The time evolution of a central cell
at the origin in $T$ and $\mu_q$ is depicted in comparison (green dashed line with green open circles).
 
It is worthwhile to connect the present discussion of the evolution path with UrQMD initial conditions to 
the results obtained with various models (however without phase transition) that are investigated in \cite{Arsene:2006vf}. The current 
findings with the chiral equation state support the main statement given there that it 
might be possible to reach the phase boundary to the QGP/chiral restoration already at moderate 
beam energies ($E_{\rm lab}\sim 5-10A$~GeV). A further discussion about the effect of different equations of state
will be presented below. It should also be noted that we do not observe a focussing of the hydrodynamical trajectories
towards the critical end point in contrast to the findings by \cite{Nonaka:2004pg,Asakawa:2005hw}. However,
see also the discussion in \cite{Scavenius:2000qd,Stephanov:2004wx}.   

\begin{figure}[t]
\centering
\includegraphics[width=9.0cm]{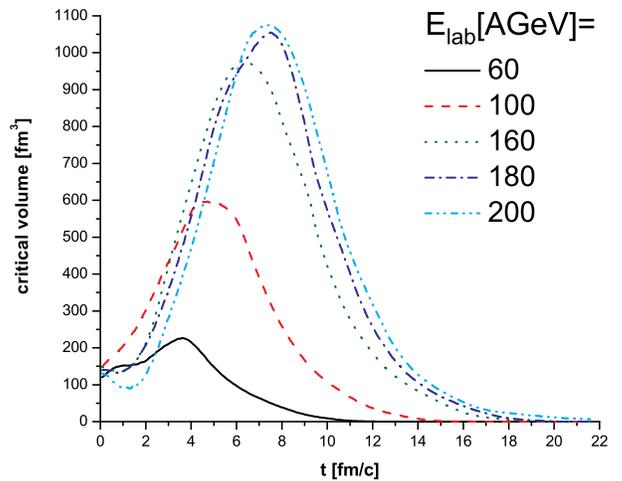}
\caption{\label{point1}
Time evolution of the critical volume for different beam energies for the CEP obtained with the chiral EoS.}
\end{figure}

Having at hand an EoS with a critical end point it is possible to explore, which fraction of the 
evolving system stays for how long close to the critical end point. The energy dependence of this exposure tine is also investigated.
Therefore, we define a 'critical volume' by adding all cells that have a temperature of
$T_{\rm CEP} \pm 10$~MeV and a chemical potential $\mu_{\rm CEP} \pm 10$~MeV for each time step. 
Fig. \ref{point1} shows the time evolution of this critical volume at various incident energies for 
the critical end point obtained by the chiral EoS. Despite the surprising fact that the maximal volume is reached for the highest beam energies ($E_{\rm lab}=160-200A~$GeV) a quite large critical volume of around $\sim 200~{\rm fm}^3$ is already predicted at lower energies of $E_{\rm lab}=60A~$GeV.   

\begin{figure}[t]
\centering
\includegraphics[width=9.0cm]{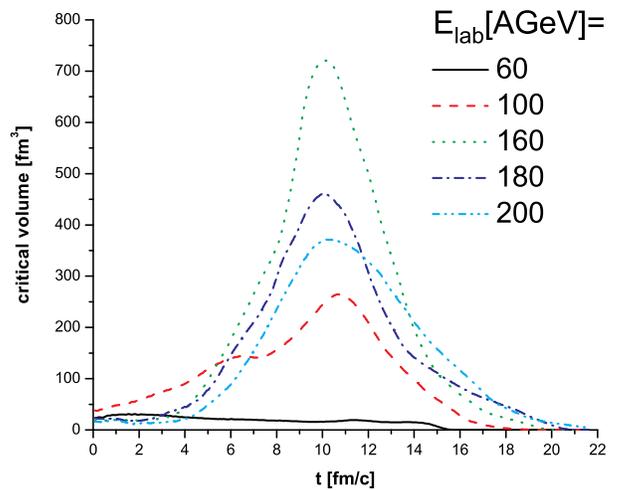}
\caption{\label{point2}
Time evolution of the critical volume for different beam energies the CEP obtained by Fodor and Katz \cite{Fodor:2004nz}.}
\end{figure}

Fig. \ref{point2} shows the critical volume for the $T_{\rm CEP}$ and $\mu_{\rm CEP}$
values obtained by lattice QCD calculations \cite{Fodor:2004nz}, however using the chiral EoS for 
the dynamics. In contrast to the chiral values in Fig. \ref{point1} the time for the maximum does not change with the energy in this case. The highest values for the critical volume are still reached at the highest beam energies, but one has to keep in mind that the critical end point for the volume calculation and the critical point in the evolution are different. The influence of different EoS on the critical volume will be discussed below.

\begin{figure}[t]
\centering
\includegraphics[width=9.0cm]{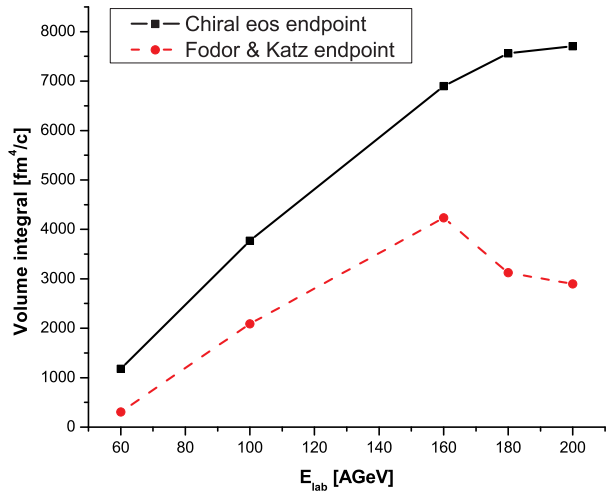}
\caption{\label{integral}
Excitation function of the critical space-time volume for the two different CEPs.}
\end{figure}

To pin down the beam energy to maximise the critical volume for the longest period of time, the total 4-volume is
obtained by integration of the critical volume over the time. 
The space-time volume is shown in Fig. \ref{integral} as a function of $E_{\rm lab}$.

Of course the actual size of the critical volume depends on the chosen $\Delta T$ and $\Delta\mu$ intervals
around $T_c$ and $\mu_c$. Also the choice of the Gaussian width of the particles in the initial condition does have
a small influence on the critical volume, since at larger $\sigma$ the distributions for  $T$ and $\mu_q$ are wider and
more 'smeared out'. However, we have checked that for the CEP obtained with the chiral 
EoS the excitation function is independent of $\sigma$.
The integrated critical volume (for the CEP obtained from the chiral model) does steadily increase 
up to $E_{\rm lab}\approx 200A~$GeV. Having the maximum critical space-time volume in the excitation 
at such an unexpectedly high beam energy is due to the way the lines of constant entropy behave at the 
phase boarder (see Fig. \ref{tmu}). With the present chiral EoS, the trajectories coming from higher 
temperatures turn right (towards lower temperatures) at the phase transition line and therefore much 
higher energies are needed as if they would turn to smaller $\mu_q$ (left) and go up the phase line. 

As a last step, we modify our chiral EoS by adding all known resonances (except the baryonic decuplett,
since it is included by means of a so called 'test'-resonance) with vacuum-masses up to 2~GeV as a free gas.
This allows to complete the baryon resonance spectrum without changing the phase structure of the EoS.
The position and type of the phase transition in the phase diagram stays unchanged, because the added resonances have
no influence on the chiral condensate $\sigma$.
\begin{figure}[t]
\centering
\includegraphics[width=9.0cm]{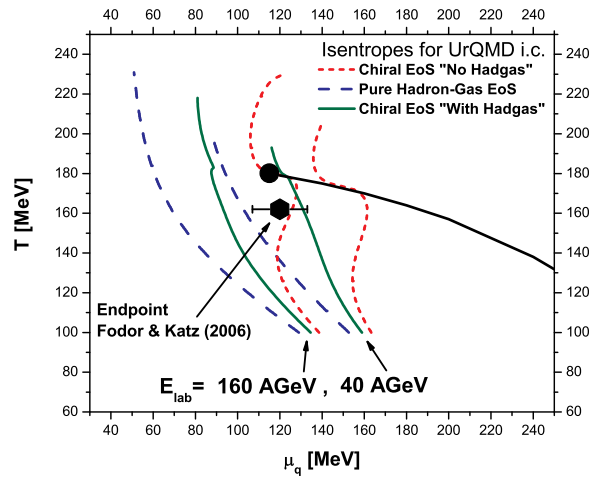}
\caption{\label{hadgas}
Isentropic expansion paths for different beam energies. The dashed lines resemble a pure hadron gas EoS. 
The short-dashed and straight lines refer to the chiral EoS without and with the completed resonance spectrum.}
\end{figure}

In Fig. \ref{hadgas} isentropic paths for the two chiral EoS (with and without completed baryon resonance spectrum)
are compared to those calculated with a free hadron-gas EoS (all particles have vacuum masses -
which is of course not consistent (and only shown for discussional purposes), since the free hadron-gas EoS 
does not have any phase transition or CEP).
As one can see, a beam energy of $40 A$~GeV is more than sufficient for an hadron gas EoS to reach the 
CEP, while the energy needed with the chiral EoS is $160 A$~GeV (without heavy resonances) and 
$E_{\rm lab}=40-60A$~GeV when heavy resonances are included.

\section{Summary}
We showed, that calculating the initial conditions of an heavy ion collision with the UrQMD model, yields very similar
$S/A$ values for a given beam energy as the simple overlap model. We then compared an isentropic expansion
scenario within a full (3+1)dimensional ideal hydrodynamic evolution with a chiral EoS including a CEP with 
constant $S/A$ lines. For the hydro evolution, the systems mean values of energy- and baryon-density
follow isentropic paths in the $\epsilon - n$ phase-diagram, while in the $T-\mu$ plane, a single cell follows 
the isentropic path, while the averaged quantities deviate from the isentropic expectation.
Most importantly it was shown, that concerning the search for the critical end point, it might not be 
sufficient to apply a free hadron gas EoS to estimate the energy needed to generate a system that, during 
its expansion, goes through the critical region. Applying different EoS (as we have done) can very much 
change predictions at what beam energy that CEP is reached.

\section*{Acknowledgements}
This work was supported by BMBF and GSI. The computational resources were provided by the Frankfurt 
Center for Scientific Computing (CSC). We would like to thank Dr. Adrian Dumitru for helpful discussions. H. Petersen thanks the Deutsche Telekom Stiftung for the scholarship and the Helmholtz Research School on Quark Matter Studies for additional suppport.


\begin{thebibliography}{99}

\bibitem{Heinz:2000bk}
  U.~W.~Heinz and M.~Jacob,
  arXiv:nucl-th/0002042.

\bibitem{Gyulassy:2004zy}
  M.~Gyulassy and L.~McLerran,
  Nucl.\ Phys.\  A {\bf 750}, 30 (2005)
  [arXiv:nucl-th/0405013].

\bibitem{Fodor:2001pe}
  Z.~Fodor and S.~D.~Katz,
  JHEP {\bf 0203}, 014 (2002)
  [arXiv:hep-lat/0106002].

\bibitem{Fodor:2007vv}
  Z.~Fodor, S.~D.~Katz and C.~Schmidt,
  JHEP {\bf 0703}, 121 (2007)
  [arXiv:hep-lat/0701022].

\bibitem{Karsch:2004wd}
  F.~Karsch,
  J.\ Phys.\ G {\bf 31}, S633 (2005)
  [arXiv:hep-lat/0412038].

\bibitem{Stephanov:1998dy}
  M.~A.~Stephanov, K.~Rajagopal and E.~V.~Shuryak,
  Phys.\ Rev.\ Lett.\  {\bf 81}, 4816 (1998)
  [arXiv:hep-ph/9806219].

\bibitem{Gazdzicki:1998vd}
  M.~Gazdzicki and M.~I.~Gorenstein,
  Acta Phys.\ Polon.\  B {\bf 30}, 2705 (1999)
  [arXiv:hep-ph/9803462].

\bibitem{Stephanov:1999zu}
  M.~A.~Stephanov, K.~Rajagopal and E.~V.~Shuryak,
  Phys.\ Rev.\  D {\bf 60}, 114028 (1999)
  [arXiv:hep-ph/9903292].

\bibitem{Bravina:1999dh}
  L.~V.~Bravina {\it et al.},
  Phys.\ Rev.\  C {\bf 60}, 024904 (1999)
  [arXiv:hep-ph/9906548].

\bibitem{Bravina:2000dk}
  L.~V.~Bravina {\it et al.},
  Phys.\ Rev.\  C {\bf 63}, 064902 (2001)
  [arXiv:hep-ph/0010172].

\bibitem{Gazdzicki:2004ef}
  M.~Gazdzicki {\it et al.}  [NA49 Collaboration],
  J.\ Phys.\ G {\bf 30}, S701 (2004)
  [arXiv:nucl-ex/0403023].

\bibitem{Arsene:2006vf}
  I.~C.~Arsene {\it et al.},
  Phys.\ Rev.\  C {\bf 75}, 034902 (2007)
  [arXiv:nucl-th/0609042].

\bibitem{Hofmann:1976dy}
  J.~Hofmann, H.~Stoecker, U.~W.~Heinz, W.~Scheid and W.~Greiner,
  Phys.\ Rev.\ Lett.\  {\bf 36}, 88 (1976).

\bibitem{Stoecker:1986ci}
  H.~Stoecker and W.~Greiner,
  Phys.\ Rept.\  {\bf 137}, 277 (1986).

\bibitem{Sorge:1998mk}
  H.~Sorge,
  Phys.\ Rev.\ Lett.\  {\bf 82}, 2048 (1999)
  [arXiv:nucl-th/9812057].

\bibitem{Ollitrault:1992bk}
  J.~Y.~Ollitrault,
  Phys.\ Rev.\  D {\bf 46}, 229 (1992).

\bibitem{Hung:1994eq}
  C.~M.~Hung and E.~V.~Shuryak,
  Phys.\ Rev.\ Lett.\  {\bf 75}, 4003 (1995)
  [arXiv:hep-ph/9412360].

\bibitem{Rischke:1996nq}
  D.~H.~Rischke,
  Nucl.\ Phys.\  A {\bf 610}, 88C (1996)
  [arXiv:nucl-th/9608024].

\bibitem{Sorge:1996pc}
  H.~Sorge,
  Phys.\ Rev.\ Lett.\  {\bf 78}, 2309 (1997)
  [arXiv:nucl-th/9610026].

\bibitem{Heiselberg:1998es}
  H.~Heiselberg and A.~M.~Levy,
  Phys.\ Rev.\  C {\bf 59}, 2716 (1999)
  [arXiv:nucl-th/9812034].

\bibitem{Soff:1999yg}
  S.~Soff, S.~A.~Bass, M.~Bleicher, H.~Stoecker and W.~Greiner,
  arXiv:nucl-th/9903061.

\bibitem{Brachmann:1999xt}
  J.~Brachmann {\it et al.},
  Phys.\ Rev.\  C {\bf 61}, 024909 (2000)
  [arXiv:nucl-th/9908010].

\bibitem{Csernai:1999nf}
  L.~P.~Csernai and D.~Rohrich,
  Phys.\ Lett.\  B {\bf 458}, 454 (1999)
  [arXiv:nucl-th/9908034].

\bibitem{Zhang:1999rs}
  B.~Zhang, M.~Gyulassy and C.~M.~Ko,
  Phys.\ Lett.\  B {\bf 455}, 45 (1999)
  [arXiv:nucl-th/9902016].

\bibitem{Kolb:2000sd}
  P.~F.~Kolb, J.~Sollfrank and U.~W.~Heinz,
  Phys.\ Rev.\  C {\bf 62}, 054909 (2000)
  [arXiv:hep-ph/0006129].

\bibitem{Bleicher:2000sx}
  M.~Bleicher and H.~Stoecker,
  Phys.\ Lett.\  B {\bf 526}, 309 (2002)
  [arXiv:hep-ph/0006147].

\bibitem{Kolb:2003dz}
  P.~F.~Kolb and U.~W.~Heinz,
  arXiv:nucl-th/0305084.

\bibitem{Stoecker:2004qu}
  H.~Stoecker,
  Nucl.\ Phys.\  A {\bf 750}, 121 (2005)
  [arXiv:nucl-th/0406018].

\bibitem{Zhu:2005qa}
  X.~l.~Zhu, M.~Bleicher and H.~Stoecker,
  Phys.\ Rev.\  C {\bf 72}, 064911 (2005)
  [arXiv:nucl-th/0509081].

\bibitem{Petersen:2006vm}
  H.~Petersen, Q.~Li, X.~Zhu and M.~Bleicher,
  Phys.\ Rev.\  C {\bf 74}, 064908 (2006)
  [arXiv:hep-ph/0608189].

\bibitem{Mishustin:1988mj}
  I.~N.~Mishustin, V.~N.~Russkikh and L.~M.~Satarov,
  Sov.\ J.\ Nucl.\ Phys.\  {\bf 48}, 454 (1988)
  [Yad.\ Fiz.\  {\bf 48}, 711 (1988)].

\bibitem{Mishustin:1989nj}
  I.~N.~Mishustin, V.~N.~Russkikh and L.~M.~Satarov,
  Nucl.\ Phys.\  A {\bf 494}, 595 (1989).

\bibitem{Katscher:1993xs}
  U.~Katscher, D.~H.~Rischke, J.~A.~Maruhn, W.~Greiner, I.~N.~Mishustin and L.~M.~Satarov,
  Z.\ Phys.\  A {\bf 346}, 209 (1993).

\bibitem{Brachmann:1997bq}
  J.~Brachmann, A.~Dumitru, J.~A.~Maruhn, H.~Stoecker, W.~Greiner and D.~H.~Rischke,
  Nucl.\ Phys.\  A {\bf 619}, 391 (1997)
  [arXiv:nucl-th/9703032].

\bibitem{Reiter:1998uq}
  M.~Reiter, A.~Dumitru, J.~Brachmann, J.~A.~Maruhn, H.~Stoecker and W.~Greiner,
  Nucl.\ Phys.\  A {\bf 643}, 99 (1998)
  [arXiv:nucl-th/9806010].

\bibitem{Bleicher:1998xi}
  M.~Bleicher {\it et al.},
  Phys.\ Rev.\  C {\bf 59}, 1844 (1999)
  [arXiv:hep-ph/9811459].

\bibitem{Brachmann:1999mp}
  J.~Brachmann, A.~Dumitru, H.~Stoecker and W.~Greiner,
  Eur.\ Phys.\ J.\  A {\bf 8}, 549 (2000)
  [arXiv:nucl-th/9912014].

\bibitem{Dumitru:2000ai}
  A.~Dumitru {\it et al.},
  Heavy Ion Phys.\  {\bf 14}, 121 (2001)
  [arXiv:nucl-th/0010107].

\bibitem{Russkikh:2003ma}
  V.~N.~Russkikh, Yu.~B.~Ivanov, E.~G.~Nikonov, W.~Norenberg and V.~D.~Toneev,
  Phys.\ Atom.\ Nucl.\  {\bf 67}, 199 (2004)
  [Yad.\ Fiz.\  {\bf 67}, 195 (2004)]
  [arXiv:nucl-th/0302029].

\bibitem{Ivanov:2005yw}
  Yu.~B.~Ivanov, V.~N.~Russkikh and V.~D.~Toneev,
  Phys.\ Rev.\  C {\bf 73}, 044904 (2006)
  [arXiv:nucl-th/0503088].

\bibitem{Toneev:2005yy}
  V.~D.~Toneev, Yu.~B.~Ivanov, E.~G.~Nikonov, W.~Norenberg and V.~N.~Russkikh,
  Phys.\ Part.\ Nucl.\ Lett.\  {\bf 2}, 288 (2005)
  [Pisma Fiz.\ Elem.\ Chast.\ Atom.\ Yadra {\bf 2N5}, 43 (2005)].

\bibitem{Baier:2006gy}
  R.~Baier and P.~Romatschke,
  Eur.\ Phys.\ J.\  C {\bf 51}, 677 (2007)
  [arXiv:nucl-th/0610108].

\bibitem{Song:2007fn}
  H.~Song and U.~W.~Heinz,
  arXiv:0709.0742 [nucl-th].

\bibitem{Papazoglou:1998vr}
  P.~Papazoglou, D.~Zschiesche, S.~Schramm, J.~Schaffner-Bielich, H.~Stoecker and W.~Greiner,
  Phys.\ Rev.\  C {\bf 59}, 411 (1999)
  [arXiv:nucl-th/9806087].

\bibitem{nucl-th/0210053}
  S.~Schramm,
  Phys.\ Lett.\  B {\bf 560}, 164 (2003)
  [arXiv:nucl-th/0210053].

\bibitem{nucl-th/0207060}
  S.~Schramm,
  Phys.\ Rev.\  C {\bf 66}, 064310 (2002)
  [arXiv:nucl-th/0207060].

\bibitem{nucl-th/0107037}
  D.~Zschiesche, S.~Schramm, H.~Stoecker and W.~Greiner,
  Phys.\ Rev.\  C {\bf 65}, 064902 (2002)
  [arXiv:nucl-th/0107037].

\bibitem{Theis:1984qc}
  J.~Theis, G.~Graebner, G.~Buchwald, J.~A.~Maruhn, W.~Greiner, H.~Stoecker and J.~Polonyi,
  Phys.\ Rev.\  D {\bf 28}, 2286 (1983).

\bibitem{nucl-th/0407117}
  D.~Zschiesche, G.~Zeeb, S.~Schramm and H.~Stoecker,
  J.\ Phys.\ G {\bf 31}, 935 (2005)
  [arXiv:nucl-th/0407117].

\bibitem{Mott}
  N.F.~Mott and H.S.W.~Massey, The Theory of Atomic Collisions,
  Oxford Press (1971), pp437.

\bibitem{Paech:2003fe}
  K.~Paech, H.~Stoecker and A.~Dumitru,
  Phys.\ Rev.\  C {\bf 68}, 044907 (2003)
  [arXiv:nucl-th/0302013].

\bibitem{Paech:2005cx}
  K.~Paech and A.~Dumitru,
  Phys.\ Lett.\  B {\bf 623}, 200 (2005)
  [arXiv:nucl-th/0504003].

\bibitem{Bass:1998ca}
  S.~A.~Bass {\it et al.},
  Prog.\ Part.\ Nucl.\ Phys.\  {\bf 41}, 255 (1998)
  [Prog.\ Part.\ Nucl.\ Phys.\  {\bf 41}, 225 (1998)]
  [arXiv:nucl-th/9803035].

\bibitem{Bleicher:1999xi}
  M.~Bleicher {\it et al.},
  J.\ Phys.\ G {\bf 25}, 1859 (1999)
  [arXiv:hep-ph/9909407].

\bibitem{Bass:1997xw}
  S.~A.~Bass {\it et al.},
  Phys.\ Rev.\ Lett.\  {\bf 81}, 4092 (1998)
  [arXiv:nucl-th/9711032].

\bibitem{Belkacem:1998gy}
  M.~Belkacem {\it et al.},
  Phys.\ Rev.\  C {\bf 58}, 1727 (1998)
  [arXiv:nucl-th/9804058].

\bibitem{Bravina:1998it}
  L.~V.~Bravina {\it et al.},
  J.\ Phys.\ G {\bf 25}, 351 (1999)
  [arXiv:nucl-th/9810036].

\bibitem{Anderlik:1998et}
  C.~Anderlik {\it et al.},
  Phys.\ Rev.\  C {\bf 59}, 3309 (1999)
  [arXiv:nucl-th/9806004].

\bibitem{Dumitru:1999sf}
  A.~Dumitru, S.~A.~Bass, M.~Bleicher, H.~Stoecker and W.~Greiner,
  Phys.\ Lett.\  B {\bf 460}, 411 (1999)
  [arXiv:nucl-th/9901046].

\bibitem{Bass:1999tu}
  S.~A.~Bass, A.~Dumitru, M.~Bleicher, L.~Bravina, E.~Zabrodin, H.~Stoecker and W.~Greiner,
  Phys.\ Rev.\  C {\bf 60}, 021902 (1999)
  [arXiv:nucl-th/9902062].

\bibitem{Magas:1999yb}
  V.~K.~Magas {\it et al.},
  Heavy Ion Phys.\  {\bf 9}, 193 (1999)
  [arXiv:nucl-th/9903045].

\bibitem{Bass:2000ib}
  S.~A.~Bass and A.~Dumitru,
  Phys.\ Rev.\  C {\bf 61}, 064909 (2000)
  [arXiv:nucl-th/0001033].

\bibitem{Teaney:2001gc}
  D.~Teaney, J.~Lauret and E.~V.~Shuryak,
  Nucl.\ Phys.\  A {\bf 698}, 479 (2002)
  [arXiv:nucl-th/0104041].

\bibitem{Teaney:2001av}
  D.~Teaney, J.~Lauret and E.~V.~Shuryak,
  arXiv:nucl-th/0110037.

\bibitem{Nonaka:2006yn}
  C.~Nonaka and S.~A.~Bass,
  Phys.\ Rev.\  C {\bf 75}, 014902 (2007)
  [arXiv:nucl-th/0607018].

\bibitem{Paiva:1996nv}
  S.~Paiva, Y.~Hama and T.~Kodama,
  Phys.\ Rev.\  C {\bf 55}, 1455 (1997).

\bibitem{Aguiar:2001ac}
  C.~E.~Aguiar, Y.~Hama, T.~Kodama and T.~Osada,
  Nucl.\ Phys.\  A {\bf 698}, 639 (2002)
  [arXiv:hep-ph/0106266].

\bibitem{Socolowski:2004hw}
  O.~.~J.~Socolowski, F.~Grassi, Y.~Hama and T.~Kodama,
  Phys.\ Rev.\ Lett.\  {\bf 93}, 182301 (2004)
  [arXiv:hep-ph/0405181].

\bibitem{Grassi:2005pm}
  F.~Grassi, Y.~Hama, O.~Socolowski and T.~Kodama,
  J.\ Phys.\ G {\bf 31}, S1041 (2005).

\bibitem{Andrade:2005tx}
  R.~Andrade, F.~Grassi, Y.~Hama, T.~Kodama, O.~.~J.~Socolowski and B.~Tavares,
  Eur.\ Phys.\ J.\  A {\bf 29}, 23 (2006)
  [arXiv:nucl-th/0511021].

\bibitem{Andrade:2006yh}
  R.~Andrade, F.~Grassi, Y.~Hama, T.~Kodama and O.~.~J.~Socolowski,
  Phys.\ Rev.\ Lett.\  {\bf 97}, 202302 (2006)
  [arXiv:nucl-th/0608067].

\bibitem{Aguiar:2007zz}
  C.~E.~Aguiar, T.~Kodama, T.~Koide and Y.~Hama,
  Braz.\ J.\ Phys.\  {\bf 37}, 95 (2007).

\bibitem{Rischke:1995ir}
  D.~H.~Rischke, S.~Bernard and J.~A.~Maruhn,
  Nucl.\ Phys.\  A {\bf 595}, 346 (1995)
  [arXiv:nucl-th/9504018].

\bibitem{Waldhauser:1992xf}
  R.~Waldhauser, D.~H.~Rischke, U.~Katscher, J.~A.~Maruhn, H.~Stoecker and W.~Greiner,
  Z.\ Phys.\  C {\bf 54}, 459 (1992).

\bibitem{Schneider:1993gd}
  V.~Schneider, U.~Katscher, D.~H.~Rischke, B.~Waldhauser, J.~A.~Maruhn and C.~D.~Munz,
  J.\ Comput.\ Phys.\  {\bf 105}, 92 (1993).

\bibitem{Rischke:1995mt}
  D.~H.~Rischke, Y.~Pursun and J.~A.~Maruhn,
  Nucl.\ Phys.\  A {\bf 595}, 383 (1995)
  [Erratum-ibid.\  A {\bf 596}, 717 (1996)]
  [arXiv:nucl-th/9504021].

\bibitem{Rischke:1995cm}
  D.~H.~Rischke and M.~Gyulassy,
  Nucl.\ Phys.\  A {\bf 597}, 701 (1996)
  [arXiv:nucl-th/9509040].

\bibitem{Gyulassy:1996br}
  M.~Gyulassy, D.~H.~Rischke and B.~Zhang,
  Nucl.\ Phys.\  A {\bf 613}, 397 (1997)
  [arXiv:nucl-th/9609030].

\bibitem{Stoecker:2007su}
  H.~Stoecker, B.~Betz and P.~Rau,
  PoS C {\bf POD2006}, 029 (2006)
  [arXiv:nucl-th/0703054].

\bibitem{Betz:2007ie}
  B.~Betz, P.~Rau and H.~Stocker,
  arXiv:0707.3942 [hep-th].

\bibitem{Nonaka:2004pg}
  C.~Nonaka and M.~Asakawa,
  Phys.\ Rev.\  C {\bf 71}, 044904 (2005)
  [arXiv:nucl-th/0410078].

\bibitem{Asakawa:2005hw}
  M.~Asakawa and C.~Nonaka,
  Nucl.\ Phys.\  A {\bf 774}, 753 (2006)
  [arXiv:nucl-th/0509091].

\bibitem{Scavenius:2000qd}
  O.~Scavenius, A.~Mocsy, I.~N.~Mishustin and D.~H.~Rischke,
  Phys.\ Rev.\  C {\bf 64}, 045202 (2001)
  [arXiv:nucl-th/0007030].

\bibitem{Stephanov:2004wx}
  M.~A.~Stephanov,
  Prog.\ Theor.\ Phys.\ Suppl.\  {\bf 153}, 139 (2004)
  [Int.\ J.\ Mod.\ Phys.\  A {\bf 20}, 4387 (2005)]
  [arXiv:hep-ph/0402115].

\bibitem{Fodor:2004nz}
  Z.~Fodor and S.~D.~Katz,
  JHEP {\bf 0404}, 050 (2004)
  [arXiv:hep-lat/0402006].

\end{thebibliography}
\end{document}